\documentclass[a4paper]{article}
\usepackage{caption}
\usepackage{subcaption}
\usepackage{INTERSPEECH2022}
\usepackage{makecell}
\usepackage{multirow}
\usepackage{graphics}
\usepackage{adjustbox}
\usepackage[bottom]{footmisc}
\title{DENT-DDSP: Data-efficient noisy speech generator using differentiable digital signal processors for explicit distortion modelling and noise-robust speech recognition}
\name{Guo Zixun$^1$, Chen Chen$^1$, Chng Eng Siong$^1$}
\address{
  $^1$Nanyang Technological University, Singapore}
 \email{zguo008@e.ntu.edu.sg}

\begin{document}

\maketitle

\begin{abstract}

The performances of automatic speech recognition (ASR) systems degrade drastically under noisy conditions. Explicit distortion modelling (EDM), as a feature compensation step, is able to enhance ASR systems under such conditions by simulating the in-domain noisy speeches from the clean counterparts. Yet, existing distortion models are either non-trainable or unexplainable and often lack controllability and generalization ability. In this paper, we propose a fully explainable and controllable model: DENT-DDSP to achieve EDM. DENT-DDSP utilizes novel differentiable digital signal processing (DDSP) components and requires only 10 seconds of training data to achieve high fidelity. The experiment shows that the simulated noisy data from DENT-DDSP achieves the highest simulation fidelity compared to other baseline models in terms of multi-scale spectral loss (MSSL). Moreover, to validate whether the data simulated by DENT-DDSP are able to replace the scarce in-domain noisy data in the noise-robust ASR tasks, several downstream ASR models with the same architecture are trained using the simulated data and the real data. The experiment shows that the model trained with the simulated noisy data from DENT-DDSP achieves similar performances to the benchmark with a 2.7\% difference in terms of word error rate (WER). The code of the model is released online\footnote[1]{https://github.com/guozixunnicolas/DENT-ddsp}.

\end{abstract}
\noindent\textbf{Index Terms}: explicit distortion modelling, DDSP, noise-robust ASR

\section{Introduction}
\label{sec:intro}

The end-to-end automatic speech recognition (ASR) systems have achieved a remarkable success, yet, the systems are prone to noisy conditions. Several datasets~\cite{rats,ATC, atc_airbus, switchboard} containing clean speeches, noisy speeches and the corresponding transcribed texts have been collected and utilized to boost ASR performances under noisy conditions. Yet, such datasets are usually scarce. Explicit distortion modelling (EDM)~\cite{survey_asr}, as an alternative, is able to simulate noisy speeches from their clean counterparts. However, distortion models with high simulation fidelity are hard to be obtained.\par

To achieve EDM, traditional methods such as parallel model combination (PMC)~\cite{gales1995model} and vector taylor series (VTS)~\cite{Minami_hmm,Sankar_hmm,Acero_vts} are widely used to estimate the distortion model before the deep learning era. An intuitive method is proposed in~\cite{MA_data_augment} where static noise within the limited collected noisy data are aggregated and added to the clean speeches. Another method~\cite{CODEC} adds pre-recorded noise to the clean speeches and passes them through pre-defined codecs to obtain the simulated noisy data. These two static methods~\cite{MA_data_augment,CODEC}, however, are untrainable thus cannot be adapted to other circumstances. Recently, GAN-based methods~\cite{chen2022noise, GAN_another} are utilized to achieve EDM and have obtained promising results. We find SimuGAN~\cite{chen2022noise} the closest match to our work. It operates directly on magnitude spectrograms and uses GAN and contrastive learning methods to distort the clean spectrograms. Yet, GAN-based methods contain large amounts of parameters making the distortion models unexplainable and uncontrollable.\par

Recently with the advent of the differentiable digital signal processing (DDSP)~\cite{EngelHGR20}, traditional digital signal processors (DSP) become differentiable and trainable and have seen success in the field of speech synthesis~\cite{speech_synth,voice_synth}. Inspired by the success of DDSP in the synthesis domain, we find DDSP a viable solution to achieve EDM. In this paper, we focus on simulating the VHF/UHF transmitted data (e.g., air traffic control). By comparing the characteristics of the VHF/UHF transmitted data with their clean counterparts, we observe that the VHF/UHF transmitted data are distorted, compressed to a fixed dynamic range, equalized and also contain colored noise. We hence find DDSP capable of achieving such exact conversion.\par



In this work, we propose a data-efficent noisy speech generator using novel DDSP components (DENT-DDSP) to simulate the VHF/UHF transmitted data. To our best knowledge, DENT-DDSP is the very first model that utilizes DDSP to achieve explicit distortion modelling. It only requires 10 seconds of parallel training data, yet, the trained distortion model achieves high fidelity simulation. As a result, the VHF/UHF transmitted data can easily be simulated from clean speeches. The simulated data can be further used for other downstream tasks (e.g., noise-robust ASR, speech enhancement).\par




More specifically, DENT-DDSP consists of two signal chains: an audio signal chain and a noise signal chain. Each signal chain consists of trainable DDSP components connected in series. The audio signal chain is able to distort, compress and equalize the clean speeches. The noise signal chain is able to transform the input white noise to the desired spectral characteristics. The outputs from the two signal chains are added to form the final simulated noisy speech. In the audio signal chain, two novel DDSP components are proposed: waveshaper and computionally-efficient dynamic range compressor (DRC). The experiment has shown that the proposed computationally-efficient DRC drastically improves the computational efficiency while maintaining the simulation fidelity.\par

Extensive experiments have shown that DENT-DDSP achieves the highest simulation fidelity among all distortion models in terms of MSSL. Moreover, it has shown strong generalization abilities over unseen training data and outperforms other GAN-based models. To further validate whether the labor-intensive way of collecting the VHF/UHF transmitted data can be replaced by simulating these data from DENT-DDSP, a downstream noise-robust ASR model trained using simulated data from DENT-DDSP is benchmarked against a same model trained with real noisy data. The experiment has shown that the model trained with simulated data from DENT-DDSP has achieved similar WER to the benchmark model with 2.7\% difference and has outperformed the best baseline model by 7.3\% in terms of WER.\par

\section{Model Architecture}\label{sec:model_architecture}

\begin{figure*}[t]
  \centering
  \includegraphics[width=.95\linewidth]{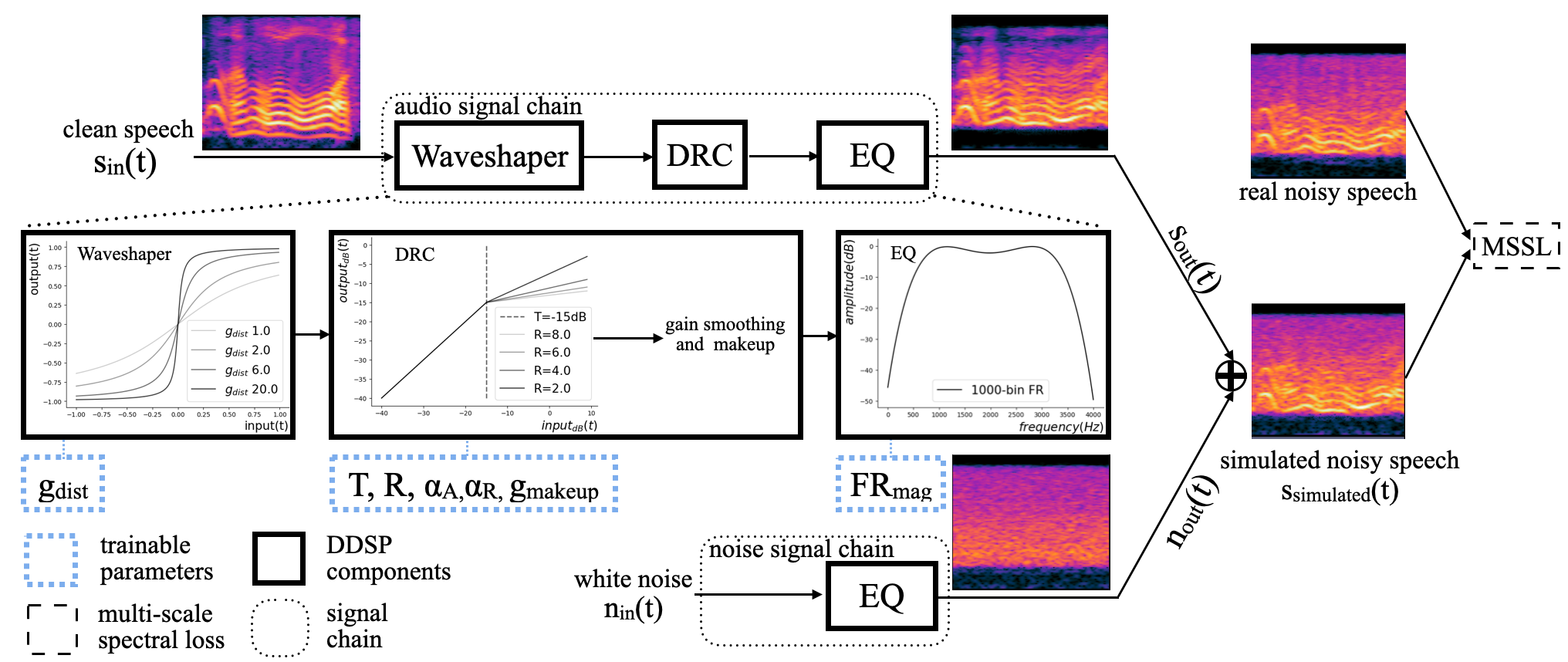}
  \caption{DENT-DDSP architecture. DRC stands for dynamic range compressor and EQ stands for equalizer. The parameters of the DDSP components are trainable and controllable.}
  \label{fig:model_arch}
\end{figure*}

The model architecture of DENT-DDSP is shown in Figure~\ref{fig:model_arch}. DENT-DDSP contains 2 parallel signal chains. The audio signal chain contains a waveshaper, a computationally-efficient DRC and an equalizer and receives clean speech: $s_{in}(t)$ as input. The noise signal chain contains an equalizer and receives white noise: $n_{in}(t)$ as input. In each signal chain, the input signal will pass through each DDSP component in series. The simulated noisy speech: $s_{simulated}(t)$ will be formed by adding the audio signal chain output: $s_{out}(t)$ and the weighted noise signal chain output:$n_{out}(t)$ following Equation~\ref{eq:final_out}. $\lambda$ is a weighting and non-trainable parameter and is set to $1$ during training. However, it can be adjusted during the generation phase to simulate noisy data with different SNRs in order to achieve data augmentation. A spectral loss function: MSSL (multi-scale spectral loss)~\cite{EngelHGR20} will be calculated between the simulated noisy speech: $s_{simulated}(t)$ and the real noisy data to update the model's parameters via backpropagation. We calculate MSSL using the following FFT sizes: 2048, 1024, 512, 256, 128, 64, to reflect the spectral distance in different spectral resolutions.\par 
\begin{equation}
  s_{simulated}(t) = s_{out}(t)+\lambda \cdot n_{out}(t)
  \label{eq:final_out}
\end{equation}
We then formulate the DDSP components in the signal chains and introduce their functionalities. For simplicity, we represent each component's input as  $x(t)$ and output as $y(t)$. $x_{dB}(t)$ and $y_{dB}(t)$ represents the input and output in the $dB$ scale.

\vspace{-0.1cm}
\subsection{Waveshaper}
The waveshaper is able to introduce a distortion effect to the input signal as shown in Equation~\ref{eq1:dist}. The distortion gain $g_{distort}$ represents the amount of distortion added to the input signal. With $g_{distort}$ close to $0$, the transfer function of a waveshaper is almost linear hence little distortion is applied. However, with an increasing $g_{distort}$, the transfer function becomes non-linear and saturates quickly. Such non-linearity introduces the distortion to the input signal. The distortion gain $g_{distort}$ is set to be a trainable parameter.

\begin{equation}
  y(t) = \frac{2}{\pi} \arctan(g_{distort} \cdot \frac{\pi}{2} \cdot x(t)) \;\;,\;\; g_{distort}>0
  \label{eq1:dist}
\end{equation}
\vspace{-0.1cm}
\subsection{Computationally-efficient DRC}
Dynamic range compressors(DRC)~\cite{queenmary_drc} are widely used in music production to limit the dynamic ranges of different audio tracks. By observation, the VHF/UHF transmitted data are compressed to a fixed dynamic range in the time domain. We hence find DRC a suitable DDSP component to achieve such limiting behaviour. A detailed survey~\cite{queenmary_drc} has introduced different kinds of DRCs and we have chosen a hard-knee DRC as our DDSP backbone. We then formulate our novel computationally-efficient DRC in Equation~\ref{eq2:drc_gain}-\ref{eq2:drc_output}.\par 
Equation~\ref{eq2:drc_gain} calculates the amplitude reduction gain $g(t)$ based on the compression threshold $T$ and the compression ratio $R$. Equation~\ref{eq2:drc_gain_downsample}-\ref{eq2:drc_gain_upsample_smooth} shows the proposed efficient implementation of gain smoothing using a novel companding operation. The amplitude reduction gain $g(t)$ is firstly downsampled via a linear interpolation by a downsampling factor $ds\_factor$ in Equation~\ref{eq2:drc_gain_downsample}. The downsampled gain function $g_{d}(t)$ is then smoothed by $\alpha_A$ and $\alpha_R$ which represents attack and release time constant respectively in Equation~\ref{eq2:drc_gain_downsample_smooth}. The smoothed gain function $g_{ds}(t)$ is then upsampled with overlapping hanning windows by an upsampling factor $us\_factor$ in Equation~\ref{eq2:drc_gain_upsample_smooth}. $ds\_factor$ and $us\_factor$ are set to be equal such that $g_{es}(t)$ will have the same shape as $g(t)$. $g_{ds}(t)$ is the main performance bottleneck in gain calculation since the operation is auto-agressive and recursive with time complexity of $O(L/ds\_factor)$ where $L$ is the total number of audio samples. Hence, the time complexity will increase linearly with a decreasing $ds\_factor$. We will later prove the companding operation in Equation~\ref{eq2:drc_gain_downsample} and \ref{eq2:drc_gain_upsample_smooth} drastically improve the efficiency while maintaining the simulation quality in Section~\ref{sec:results}. Finally, the output $y_{dB}(t)$ is obtained in Equation~\ref{eq2:drc_output} with an additional makeup gain $g_{makeup}$. The following parameters are trainable: $T$, $R$, $\alpha_A$, $\alpha_R$, $g_{makeup}$. 


%
\begin{equation}
  g(t) =
    \begin{cases}
      0 & \text{$x_{dB}(t)\le T$}\\     
      \displaystyle \frac{x_{dB}(t) - T }{R} & \text{$x_{dB}(t)> T$}\\
    \end{cases}   
    \label{eq2:drc_gain}
\end{equation}
\begin{equation}
  g_d(t) =downsample(g(t), ds\_factor)
    \label{eq2:drc_gain_downsample}
\end{equation}
%
\begin{equation}
g_{ds}(t) =
    \begin{cases}
      \alpha_Ag_{ds}(t-1) + (1-\alpha_A)g_d(t) & \text{$g_d(t)> g_{ds}(t-1)$}\\ 
      \alpha_Rg_{ds}(t-1) + (1-\alpha_R)g_d(t) & \text{$g_d(t)\le g_{ds}(t-1)$}\\    
    \end{cases}   
\label{eq2:drc_gain_downsample_smooth}
\end{equation}
\begin{equation}
  g_{es}(t) =upsample(g_{ds}(t), us\_factor)
    \label{eq2:drc_gain_upsample_smooth}
\end{equation}
\begin{equation}
  y_{dB}(t) = x_{dB}(t) + g_{es}(t) +g_{makeup}
 \label{eq2:drc_output}
\end{equation}

\subsection{Equalizer(EQ)}
By observation, the VHF/UHF transmitted data are band-limited and equalized in the frequency domain. To achieve this, we adopt the equalizer (EQ) implementation from~\cite{EngelHGR20} where the EQ is implemented using a linear time invariant FIR filter. For each EQ, the trainable parameters are set to be the magnitude of the filter frequency response: $FR_{mag}$ and the number of the frequency bins are set to be 1000. Additionally, for the EQ in the noise signal chain, the noise amplitude is made trainable in order to adjust the volume of the filtered noise. 


\section{Experiments and results}
\label{sec:experiment_setup}
\subsection{Dataset}

The Robust automatic transcription of speech (RATS) project \cite{rats} has collected a parallel dataset which contains clean speeches, the corresponding VHF/UHF transmitted noisy speeches and the transcribed texts. To obtain the dataset, pre-recorded conversational speeches are broadcasted over 8 radio channels and the transmitted noisy audio are captured concurrently. We select Channel A from the RATs dataset which contains 57.4 hours of data for training and testing. To address the data scarcity problem mentioned in Section~\ref{sec:intro}, only less than 60 seconds of parallel audio are selected as training data. During training, the data are batched into 1-second chunks.




\subsection{Experiment setup}
\subsubsection{Training data selection and model comparison}
\label{sec:baseline}
To obtain the best performing and most efficient distortion model, several DENT-DDSP models are trained and compared using different amounts of parallel training data from 10 to 60 seconds with various speech to total ratio: $s2t$. $s2t$ is defined as the duration of active speech over the total duration in each 1-second chunk. Active speech duration is calculated based on the clean audio energy in a sliding window with a threshold of $50dB$. By analyzing the dataset, with $s2t<0.4$, the data contains mostly non-speech audio or silence. Hence, we only select data with $0.4 \leq s2t<1.0$. To acquire the desired training data, the 1-second data chunks with the desired $s2t$ are randomly chosen and aggregated to the desired total duration. To evaluate the simulation fidelity, testing clean speeches are input to the trained distortion model and the MSSL\cite{EngelHGR20} is calculated between the real and simulated noisy data. We compare the proposed DENT-DDSP with the following baseline distortion models in terms of simulation fidelity: 
\begin{itemize}
    \item Clean augment model\cite{MA_data_augment}: simulated noisy speeches are obtained by adding the aggregated stationary noise from RATs Channel A to the clean speeches. 
    \vspace{-1mm}
    \item G.726 augment model\cite{CODEC}: The clean speeches are first passed through the G.726 codec. The aggregated stationary noise from RATs Channel A is added subsequently to the codec outputs.     \vspace{-1mm}
    \item Codec2 augment model\cite{CODEC}: Codec choice in the G.726 augment model is replaced to Codec2 with 700 bits/s.    \vspace{-1mm} 
    \item SimuGAN\cite{chen2022noise}: simulated noisy speeches are obtained by feeding the clean speeches to a trained SimuGAN.     \vspace{-1mm}
\end{itemize}




\subsubsection{Noise-robust automatic speech recognition (ASR)}
A downstream noise-robust ASR task is performed to validate the effectiveness of DENT-DDSP. The 57.4-hour parallel data from Channel A are split into 3 folds~\cite{chen2022self}: 44.3-hour data for training; 4.9-hour data for validation and 8.2-hour data for testing. We adopted the conformer-based~\cite{conformer} dual-path ASR system from~\cite{chen2022noise} with 12 conformer layers in the encoder and 6 transformer layers in the decoder. Such dual-path ASR architecture is specifically designed for noise-robust ASR~\cite{hu2022interactive, hu2022dual}. We pre-train a language model~\cite{lm} with 2 RNN layers using the existing transcribed texts and utilize it during text decoding. To benchmark the performance, a benchmark model is firstly trained with real noisy data. To address the data scarcity problem aforementioned in the real world, the benchmark model only incorporates the single-path setting in~\cite{chen2022noise} where only real noisy data and the transcribed text are input to the network . Another set of model is trained using the simulated noisy data from various distortion models using the dual-path setting in~\cite{chen2022noise}. These sets of models will be tested on a common test set consists of real noisy data and the performances in terms of WER will be compared. 
\vspace{-0.1cm}
\subsection{Results}
\label{sec:results}

\subsubsection{Effect of different amounts of training data with various $s2t$ and model comparison}
\begin{figure}[h]
  \centering
  \includegraphics[width=.85\linewidth]{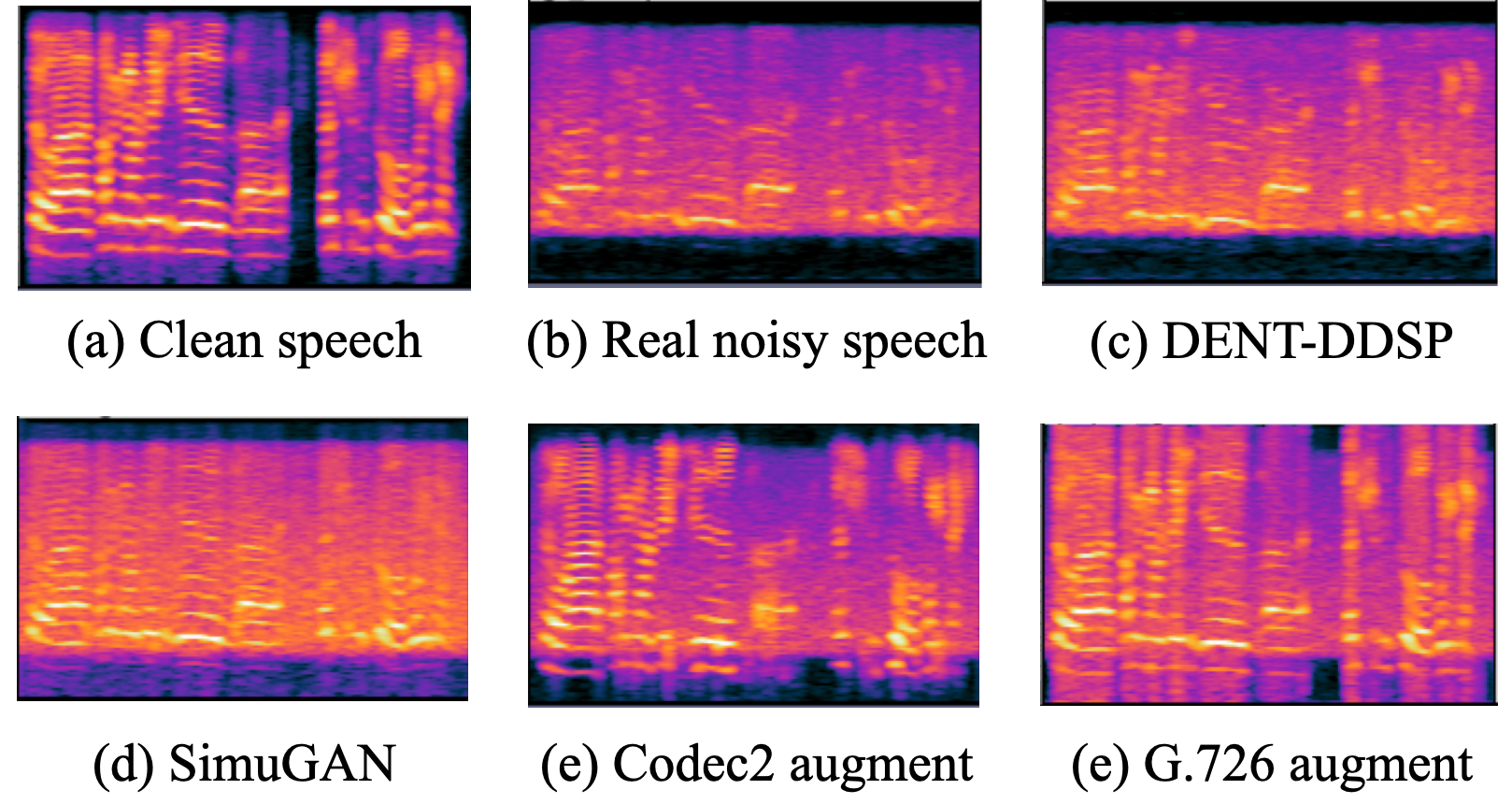}
  \caption{Magnitude spectrogram comparison between clean audio, audio simulated from various distortion models and the real noisy data.}
  \label{fig:freq_comp_dent_simugan}
\end{figure}

Table~\ref{tab:training_dur_s2t} shows the MSSL and between the real and simulated audio using different amounts of training data with various $s2t$. In general, MSSL decreases with fewer amounts of training data and increasing $s2t$. The best DENT-DDSP model is hence obtained using 10 seconds of training data with $0.8 \leq s2t<1.0$. Comparing DENT-DDSP with other distortion models, we observe that DENT-DDSP is trainable, contains much fewer trainable parameters and requires much fewer amount of training data. Moreover, noisy audio simulated by DENT-DDSP has the lowest MSSL which indicates DENT-DDSP has the highest simulation fidelity among all baseline models. Readers are strongly encouraged to listen to the simulated noisy speeches and compare them with the real noisy data from\footnote[2]{\label{foot2}https://guozixunnicolas.github.io/DENT-DDSP-demo/}.
To visualize the simulation fidelity, example spectrograms of the clean audio, audio simulated from various distortion models and the real noisy data are plotted in Fig~\ref{fig:freq_comp_dent_simugan}.


\begin{table}[t]
  \caption{Simulation quality of DENT-DDSP using different amounts of training data with various s2t and model comparison}
  \label{tab:training_dur_s2t}
  \centering
\begin{adjustbox}{width=.8\columnwidth,center}
  \begin{tabular}{lllll}
    \toprule
    \textbf{model}&\makecell[l]{\textbf{no. of} \\\textbf{trainable} \\\textbf{param.}}   &\textbf{s2t}   &\makecell[l]{\textbf{amount of} \\\textbf{training data}}         & \textbf{MSSL}          \\
    \midrule
    \multirow{12}{*}{DENT-DDSP (ours)} &\multirow{12}{*}{~2k}  & $[0.8, 1,0)$    &60-sec parallel  &0.184     \\
                                                        &&$[0.6, 0.8)$     &60-sec parallel  &0.187  \\ 
                                                        &&$[0.4, 0.6)$     &60-sec parallel  &0.186   \\ 
                                                        
                                                        &&$[0.8, 1,0)$     &40-sec parallel  &0.177   \\
                                                        &&$[0.6, 0.8)$     &40-sec parallel  &0.187    \\ 
                                                        &&$[0.4, 0.6)$     &40-sec parallel  &0.188    \\ 
                                                        
                                                        &&$[0.8, 1,0)$     &20-sec parallel  &0.170 \\
                                                        &&$[0.6, 0.8)$     &20-sec parallel  &0.189  \\ 
                                                        
                                                        &&$[0.4, 0.6)$     &20-sec parallel  &0.189   \\ 
                                                    
                                                        &&$[0.8, 1,0)$     &10-sec parallel &\textbf{0.170}  \\
                                                        &&$[0.6, 0.8)$     &10-sec parallel  &0.181   \\ 
                                                        &&$[0.4, 0.6)$     &10-sec parallel  &0.189  \\ 
    \midrule
    SimuGAN  &~14M   &-&10-min unparallel& 0.173    \\
    \midrule
    Clean augment  &non-trainable   &- & -&   0.192 \\
    \midrule
    G.726 augment  &non-trainable &-&-&     0.197\\
    \midrule
    Codec2 augment &non-trainable   &-&-&    0.197\\
    \bottomrule
  \end{tabular}
 \end{adjustbox}

\end{table}

\subsubsection{Effectiveness of the companding operation in the proposed computationally-efficient DRC}
Table~\ref{tab:companding} shows the effectiveness of the proposed companding operation in the novel computationally-efficient DRC. With decreasing $ds\_factor$s using the best DENT-DDSP obtained, the training time increases linearly, however, without obvious improvements on the MSSL. This justifies the effectiveness of the companding operation which is able to increase computational efficiency while maintaining the generation quality. 
\begin{table}
  \caption{Training time and MSSL comparison among DENT-DDSPs with different $ds\_factor$s}
  \label{tab:companding}
  \centering
  \begin{adjustbox}{width=.45\columnwidth,center}

  \begin{tabular}{lll}
    \toprule

    \textbf{ds factor}          & \textbf{training time}   & \textbf{MSSL}  \\   
    \midrule
    $16$     &~7min &0.170    \\
    $8$     &~13.5min &0.169   \\ 
    $4$      &~25.5min & 0.170    \\ 
    $2$      &~60.2min & 0.169  \\

    \bottomrule
  \end{tabular}
 \end{adjustbox}
\end{table}

\subsubsection{Generalization ability}
\begin{figure}
  \centering
  \includegraphics[width=\linewidth]{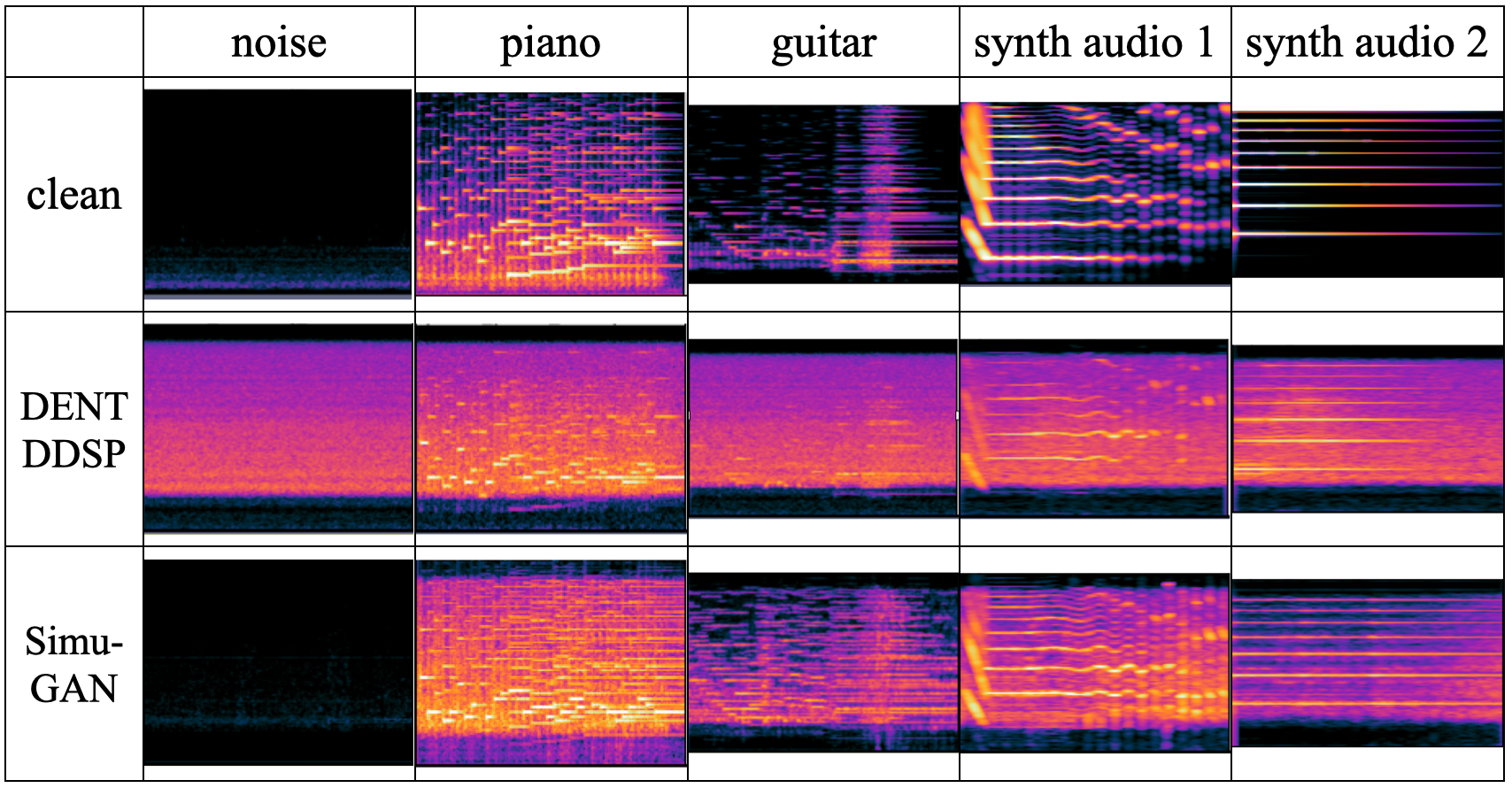}
  \caption{Noisy data simulated by DENT-DDSP and SimuGAN using unseen non-speech data}
  \label{fig:test_generalization}
\vspace{-0.1in}
\end{figure}

To reflect the generalization ability of the trainable distortion models: DENT-DDSP and SimuGAN over different types of audio, non-speech audio are used to test the distortion models. The non-speech audio (available online\footref{foot2}) include ambient noise, guitar and piano music and synthesized audio. The spectrograms of the testing audio and the simulated noisy audio from the two models are shown in Figure~\ref{fig:test_generalization}. By observation, data simulated by DENT-DDSP have consistent noise spectrums and the audio are distorted to the desired spectral behaviour: the spectrogram becomes blurred with certain frequencies being dampened. Yet, data simulated by SimuGAN fail to produce such consistent distortion behaviour. Moreover, data simulated by SimuGAN either contain artefacts or become unintelligible. Hence, DENT-DDSP shows stronger generalization ability compared to SimuGAN due to the use of explainable DDSP components.

\subsubsection{Benchmarking the noise-robust ASR task}

To compare against the benchmark model, several additional models are obtained as follows: Data simulated by DENT augment model is obtained by setting $\lambda=0.79, 1.26$ in Equation~\ref{eq:final_out} in a trained DENT-DDSP model so that the noise gain of the $n_{out}(t)$ become $\pm2dB$; Additionally, a clean model is trained only using clean speeches. Table~\ref{tab:asr_results} shows the WER of these models tested on a common test set consists of real noisy data. By observation, the DENT augment model achieves the closest WER compared to the benchmark model with a difference of 2.7\% and outperforms the best baseline model: SimuGAN by 7.3\%. This proves that DENT-DDSP is able to simulate noisy data with similar characteristics to the real noisy data and offers an alternative to simulate the VHF/UHF transmitted data thus greatly reducing the effort to collect these data manually on a large scale.





\begin{table}[h]
  \caption{WER comparison using same ASR model with simulated data from different distortion models and real noisy data}
  \label{tab:asr_results}
  \centering
  \begin{adjustbox}{width=.65\columnwidth,center}

  \begin{tabular}{lll}
    \toprule
    \textbf{model}    & \makecell[l]{\textbf{noisy speech source} } & \textbf{WER} \\
    \midrule

    \makecell[l]{Clean}    & \makecell[l]{-}&93.4\%\\
    \midrule

    \makecell[l]{Clean \\augment}    & \makecell[l]{clean speech + \\aggregated noise}&73.8\%\\

    \midrule
    \makecell[l]{G.726 \\augment}    & \makecell[l]{g726 codec speech + \\aggregated noise}& 75.6\%\\
    \midrule

    \makecell[l]{Codec2 \\augment}    & \makecell[l]{codec2 speech + \\aggregated noise}&83.2\%\\
    
    \midrule

    SimuGAN                     &\makecell[l]{SimuGAN simulated}  &65.9\%\\
    \midrule

    DENT-DDSP (ours)                     &\makecell[l]{DENT simulated}  &66.4\%\\
    \midrule

    \makecell[l]{DENT augment (ours)}     &\makecell[l]{DENT simulated with \\ noise gain $=\pm2dB$} & \textbf{58.6\%}  \\
    \midrule
    \midrule
    \makecell[l]{Benchmark}        & \makecell[l]{real noisy data} & 55.9\%\\


    \bottomrule
  \end{tabular}
 \end{adjustbox}
\end{table}

\vspace{-0.3cm}
\section{Conclusion}
\label{sec:conclusion}
We propose a fully explainable and controllable model DENT-DDSP based on novel DDSP components which only requires 10 seconds of training data to achieve explicit distortion modelling. Besides the existing DDSP components, we propose 2 novel DDSP components: waveshaper and computationally-efficient DRC which can be easily integrated to other DDSP models. Experiments have shown that data simulated by DENT-DDSP achieve the highest simulation fidelity among all distortion models. Moreover, the noise-robust ASR system trained with the data simulated by DENT-DDSP achieves similar WER compared to the benchmark trained with real noisy data with a 2.7\% difference in terms of WER. It has also achieved 7.3\% WER improvement over the best baseline distortion model.  


\bibliographystyle{IEEEtran}
\bibliography{mybib}

\begin{thebibliography}{10}
\providecommand{\url}[1]{#1}
\csname url@samestyle\endcsname
\providecommand{\newblock}{\relax}
\providecommand{\bibinfo}[2]{#2}
\providecommand{\BIBentrySTDinterwordspacing}{\spaceskip=0pt\relax}
\providecommand{\BIBentryALTinterwordstretchfactor}{4}
\providecommand{\BIBentryALTinterwordspacing}{\spaceskip=\fontdimen2\font plus
\BIBentryALTinterwordstretchfactor\fontdimen3\font minus
  \fontdimen4\font\relax}
\providecommand{\BIBforeignlanguage}[2]{{%
\expandafter\ifx\csname l@#1\endcsname\relax
\typeout{** WARNING: IEEEtran.bst: No hyphenation pattern has been}%
\typeout{** loaded for the language `#1'. Using the pattern for}%
\typeout{** the default language instead.}%
\else
\language=\csname l@#1\endcsname
\fi
#2}}
\providecommand{\BIBdecl}{\relax}
\BIBdecl

\bibitem{rats}
D.~Graff, K.~Walker, S.~Strassel, X.~Ma, K.~Jones, and A.~Sawyer, ``The {RATS}
  collection: Supporting {HLT} research with degraded audio data,'' in
  \emph{Proc. of the 9th Int. Conf. on Lang. Resour. and Eval. ({LREC})}, 2014,
  pp. 1970--1977.

\bibitem{ATC}
S.~Badrinath and H.~Balakrishnan, ``Automatic speech recognition for air
  traffic control communications,'' \emph{Transportation Research Record}, vol.
  2676, no.~1, pp. 798--810, 2022.

\bibitem{atc_airbus}
T.~Pellegrini, J.~Farinas, E.~Delpech, and F.~Lancelot, ``The airbus air
  traffic control speech recognition 2018 challenge: towards atc automatic
  transcription and call sign detection,'' in \emph{Proc. Int. Speech Commun.
  Assoc.({Interspeech})}, 2018, pp. 2993--2997.

\bibitem{switchboard}
J.~Godfrey, E.~Holliman, and J.~McDaniel, ``Switchboard: telephone speech
  corpus for research and development,'' in \emph{Proc. IEEE Int. Conf. on
  Acoustics, Speech and Signal Processing ({ICASSP})}, vol.~1, 1992, pp.
  517--520 vol.1.

\bibitem{survey_asr}
J.~Li, L.~Deng, Y.~Gong, and R.~Haeb-Umbach, ``An overview of noise-robust
  automatic speech recognition,'' \emph{IEEE ACM Trans. on Audio, Speech, and
  Lang. Process.}, vol.~22, no.~4, pp. 745--777, 2014.

\bibitem{gales1995model}
M.~J.~F. Gales, ``Model-based techniques for noise robust speech recognition,''
  in \emph{Ph.D. thesis, University of Cambridge}, 1995.

\bibitem{Minami_hmm}
Y.~Minami and S.~Furui, ``A maximum likelihood procedure for a universal
  adaptation method based on hmm composition,'' in \emph{Proc. IEEE Int. Conf.
  on Acoustics, Speech and Signal Processing ({ICASSP})}, vol.~1, 1995, pp.
  129--132.

\bibitem{Sankar_hmm}
A.~Sankar and C.-H. Lee, ``Robust speech recognition based on stochastic
  matching,'' in \emph{Proc. IEEE Int. Conf. on Acoustics, Speech and Signal
  Processing ({ICASSP})}, vol.~1, 1995, pp. 121--124.

\bibitem{Acero_vts}
A.~Acero, l.~Deng, T.~Kristjansson, and J.~Zhang, ``Hmm adaptation using vector
  taylor series for noisy speech recognition,'' 2000, pp. 869--872.

\bibitem{MA_data_augment}
D.~Ma, G.~Li, H.~Xu, and E.~S. Chng, ``Improving code-switching speech
  recognition with data augmentation and system combination,'' in \emph{Proc.
  Asia-Pacific Signal and Inf. Process. Assoc. Annu. Summit and Conf. ({APSIPA
  ASC})}, 2019, pp. 1308--1312.

\bibitem{CODEC}
M.~Ferràs, S.~Madikeri, P.~Motlicek, S.~Dey, and H.~Bourlard, ``A large-scale
  open-source acoustic simulator for speaker recognition,'' \emph{IEEE Signal
  Processing Letters}, vol.~23, no.~4, pp. 527--531, 2016.

\bibitem{chen2022noise}
C.~Chen, N.~Hou, Y.~Hu, S.~Shirol, and E.~S. Chng, ``Noise-robust speech
  recognition with 10 minutes unparalleled in-domain data,'' in \emph{Proc.
  IEEE Int. Conf. on Acoustics, Speech and Signal Processing ({ICASSP})}, 2022,
  pp. 4298--4302.

\bibitem{GAN_another}
H.~Hu, T.~Tan, and Y.~Qian, ``Generative adversarial networks based data
  augmentation for noise robust speech recognition,'' in \emph{Proc. IEEE Int.
  Conf. on Acoustics, Speech and Signal Processing ({ICASSP})}, 2018, pp.
  5044--5048.

\bibitem{EngelHGR20}
J.~H. Engel, L.~Hantrakul, C.~Gu, and A.~Roberts, ``{DDSP:} differentiable
  digital signal processing,'' in \emph{in Proc. Int. Conf. on Learn.
  Representations({ICLR})}, 2020.

\bibitem{speech_synth}
\BIBentryALTinterwordspacing
G.~Fabbro, V.~Golkov, T.~Kemp, and D.~Cremers, ``Speech synthesis and control
  using differentiable dsp,'' 2020. [Online]. Available:
  \url{https://arxiv.org/abs/2010.15084}
\BIBentrySTDinterwordspacing

\bibitem{voice_synth}
J.~Alonso and C.~Erkut, ``Latent space explorations of singing voice synthesis
  using ddsp,'' in \emph{Proc. of 18th Sound and Music Computing Conf.}, 2021.

\bibitem{queenmary_drc}
D.~Giannoulis, M.~Massberg, and J.~Reiss, ``Digital dynamic range compressor
  design—a tutorial and analysis,'' \emph{J. of the Audio Eng. Soc.({AES})},
  vol.~60, pp. 399--408, 2012.

\bibitem{chen2022self}
C.~Chen, Y.~Hu, N.~Hou, X.~Qi, H.~Zou, and E.~S. Chng, ``Self-critical sequence
  training for automatic speech recognition,'' in \emph{Proc. IEEE Int. Conf.
  on Acoustics, Speech and Signal Processing ({ICASSP})}, 2022, pp. 3688--3692.

\bibitem{conformer}
A.~Gulati, J.~Qin, C.~Chiu, N.~Parmar, Y.~Zhang, J.~Yu, W.~Han, S.~Wang,
  Z.~Zhang, Y.~Wu, and R.~Pang, ``Conformer: Convolution-augmented transformer
  for speech recognition,'' in \emph{Proc. Int. Speech Commun.
  Assoc.({Interspeech})}, 2020, pp. 5036--5040.

\bibitem{hu2022interactive}
Y.~Hu, N.~Hou, C.~Chen, and E.~S. Chng, ``Interactive feature fusion for
  end-to-end noise-robust speech recognition,'' in \emph{Proc. IEEE Int. Conf.
  on Acoustics, Speech and Signal Processing ({ICASSP})}, 2022, pp. 6292--6296.

\bibitem{hu2022dual}
Y.~Hu, N.~Hou, C.~Chen, and E.~Chng, ``Dual-path style learning for end-to-end
  noise-robust speech recognition,'' in \emph{preprint arXiv2203.14838}, 2022.

\bibitem{lm}
S.~Karita, N.~Chen, T.~Hayashi, T.~Hori, H.~Inaguma, Z.~Jiang, M.~Someki,
  N.~E.~Y. Soplin, R.~Yamamoto, X.~Wang, S.~Watanabe, T.~Yoshimura, and
  W.~Zhang, ``A comparative study on transformer vs rnn in speech
  applications,'' \emph{IEEE Automatic Speech Recognition and Understanding
  Workshop (ASRU)}, Dec 2019.

\end{thebibliography}

\end{document}